\documentclass[twocolumn]{aastex631}
\usepackage{amsmath}
\usepackage{natbib}
\usepackage{graphicx}
\usepackage{url}

\usepackage{verbatim}

\shorttitle{Water Ice in 114--426}
\shortauthors{Ballering et al.}

\begin{document}

\title{Water Ice in the Edge-On Orion Silhouette Disk 114--426 from JWST NIRCam Images}

\author[0000-0002-4276-3730]{Nicholas P. Ballering}
\affiliation{Space Science Institute, Boulder, CO 80301, USA}
\affiliation{Department of Astronomy, University of Virginia, Charlottesville, VA 22904, USA}

\author[0000-0003-2076-8001]{L. Ilsedore Cleeves}
\affiliation{Department of Astronomy, University of Virginia, Charlottesville, VA 22904, USA}
\affiliation{Department of Chemistry, University of Virginia, Charlottesville, VA 22904, USA}

\author[0000-0001-9857-1853]{Ryan D. Boyden}
\affiliation{Department of Astronomy, University of Virginia, Charlottesville, VA 22904, USA}
\affiliation{Space Science Institute, Boulder, CO 80301, USA}

\author[0000-0002-1452-5268]{Mark J. McCaughrean}
\affiliation{Max-Planck-Institut f\"ur Astronomie, K\"onigstuhl~17, 69117 Heidelberg, DE}

\author[0000-0002-0477-6047]{Rachel E. Gross}
\affiliation{Department of Chemistry, University of Virginia, Charlottesville, VA 22904, USA}

\author[0000-0001-7724-815X]{Samuel G. Pearson}
\affiliation{European Space Research and Technology Centre (ESTEC), European Space Agency, Postbus 299, 2200AG Noordwijk, NL}

\correspondingauthor{Nicholas P. Ballering}
\email{nballering@spacescience.org}

\begin{abstract}
We examine images of the protoplanetary disk 114--426 with JWST/NIRCam in 12 bands. This large disk is oriented edge on with a dark midplane flanked by lobes of scattered light. The outer edges of the midplane are seen in silhouette against the Orion Nebula, providing a unique opportunity to study planet-forming material in absorption. We discover a dip in the scattered light of the disk at 3\,$\micron$ -- compelling evidence for the presence of water ice. The 3\,$\micron$ dip is also seen in the silhouette of the disk, where we quantify the ice abundance with models of pure absorption and avoid the complications of disk scattering effects. We find grain ice-to-refractory mass ratios of up to $\sim$0.2, maximum grain sizes of 0.25--5\,$\micron$, and a total dust plus ice mass of 0.46\,$M_\oplus$ in the silhouette region. We also discover excess absorption in the NIRCam bands that include the Pa$\alpha$ line, suggesting there may be excited atomic hydrogen in the disk. Examining the morphology of the scattered light lobes reveals that they are laterally offset from each other and exhibit a brightness asymmetry that flips with wavelength -- both evidence for a tilted inner disk in this system.
\end{abstract}

\section{INTRODUCTION}
\label{sec:introduction}
Water is vital to life as we know it. During the formation of the solar system, much of the water resided as ice mantles on dust grains beyond the snow line of the protosolar disk. Icy mantles increase the sticking efficiency of dust grains \citep{gundlach2015_stickyice}, facilitating their growth to larger pebbles and planetesimals in which the ice becomes sequestered. Theories suggest water may have been delivered to the inner solar system by inward-drifting pebbles or scattered planetesimals. Today, much water remains in the outer solar system, frozen in bodies like Kuiper Belt objects and icy moons.

To understand the evolution of water ice in our solar system and its role in the evolution of other planetary systems, it is essential to observe ice throughout the star and planet formation process. In general, icy grains can be detected by measuring distinct IR spectral features due to vibrational modes in the ice. Water ice has a prominent feature at 3 $\micron$\footnote{In addition to water, NH$_3$ and CH$_3$OH ices can also contribute to the 3 $\micron$ feature.} that has been detected in absorption through molecular clouds \citep[e.g.,][]{mcclure2023_iceageclouds}, dense cores \citep[e.g.,][]{boogert2011_icecores}, and protostellar envelopes \citep[e.g.,][]{yang2022_CORINOSI}.

\begin{figure*}
\epsscale{1.16}
\plotone{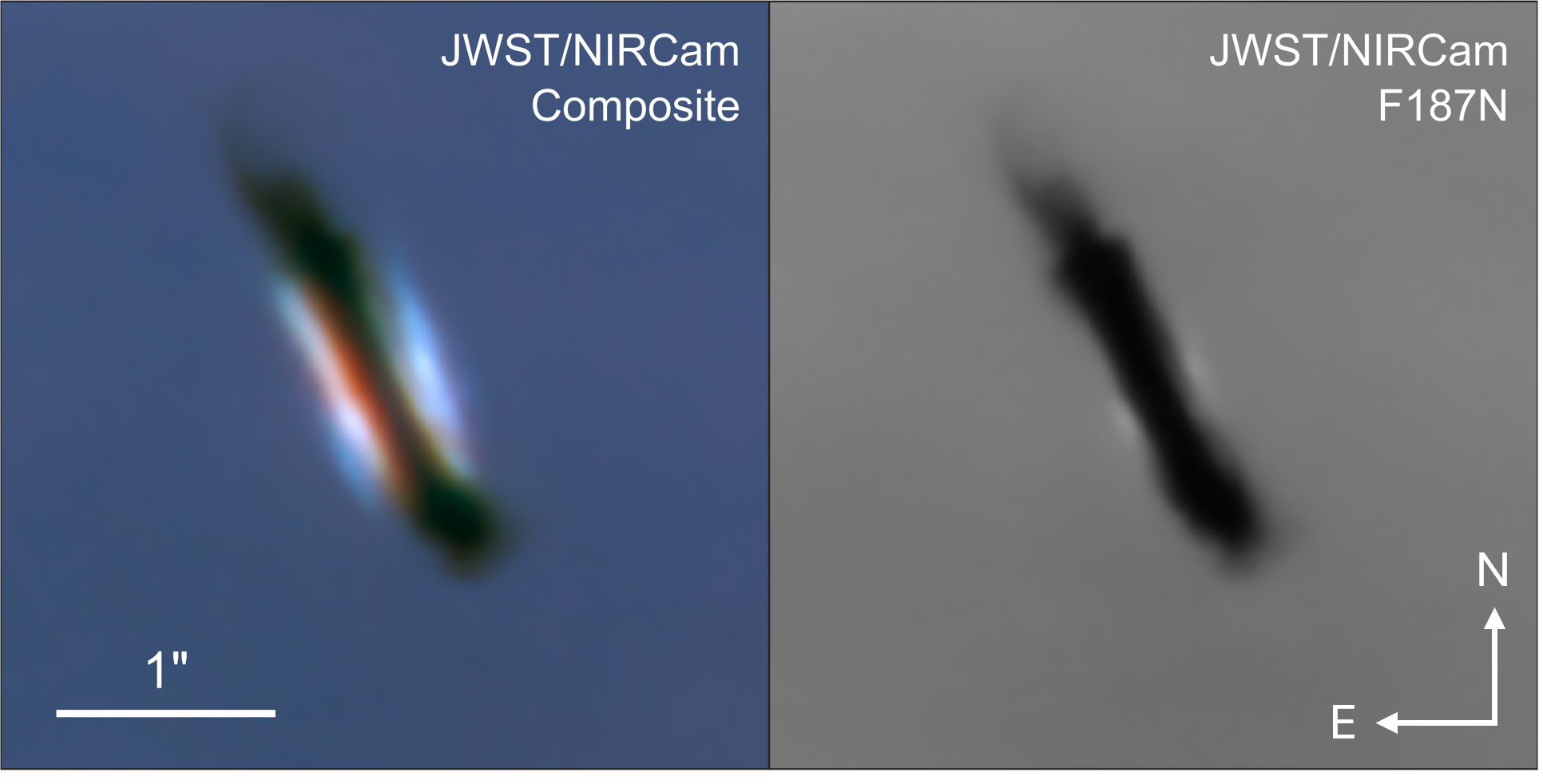}
\caption{Left: color composite image of 114--426 using the F115W and F140M (both blue), F162M (green), F182M (yellow-green), and F277W (red) bands. Each of the four SW channel images (F115W, F140M, F162M, and F182M) were interpolated onto a $3\times 3$ pixel subgrid (i.e.\ roughly 10 mas pixel$^{-1}$) to allow precise alignment, and the LW channel image (F277W) was resampled to match. Right: an image of 114--426 in the F187N band, again interpolated onto a $3\times 3$ pixel subgrid, showing the silhouette disk in maximum contrast against the Pa$\alpha$ line. Each panel is 3$\farcs$5 $\times$ 3$\farcs$5 with north up and east left. The intensity of each image was scaled logarithmically to compress the dynamic range and enhance fainter details.}
\label{fig:highres}
\end{figure*}

The 3\,$\micron$ feature has been observed toward a number of edge-on protoplanetary disks \citep{pontoppidan2005_iceCRBR,terada2007_waterice,mccabe2011_HKtau,sturm2023_HH48ERS}. In these systems, the scattered light from the star and warm inner disk serves as the background continuum on which the ice features are imprinted. The scattering of radiation through the disk complicates the interpretation of these observations \citep{ballering2021_diskices,sturm2023_HH48modelII} such that the column density of ice cannot be directly inferred from the depth of the absorption feature, as is possible in the case of pure absorption against a background source.

The disk 114--426 provides a unique opportunity to study planet-forming material in absorption. 114--426, named using the coordinate-based convention introduced by \citet{odell1994_HSTONC}, resides in the dense Trapezium Cluster, itself part of the wider Orion Nebula Cluster (ONC). It is located to the southwest of the Trapezium OB stars. The disk has a major axis of $2\farcs7$, corresponding to a diameter of 1053\,au at a distance of 390\,pc \citep{maiz-apellaniz2022_GaiaOB}. It is oriented edge on, as evident by the dark lane of the disk midplane that occults the central star, with surrounding lobes of scattered light. Additionally, the outer ansae of the dark lane extend beyond the radial extent of the scattered light and are seen in silhouette against the emission of the background Orion Nebula. This provides a region where the disk can be studied in pure absorption.

114--426 has been the subject of several detailed imaging studies at optical and near-IR wavelengths, primarily with the Hubble Space Telescope \citep[HST;][]{mccaughrean1996AJ_silhouettes,mccaughrean1998_114-426,bally2000_ONC,throop2001_114-426,shuping2003_114-426,miotello2012_114-426}. It was realized that the outer ansae of the disk are somewhat transparent to the background nebula at optical wavelengths. Examining the variation in optical depth with wavelength provides a means to measure the dust column density and grain sizes. The most recent analysis used absorption measurements in five visible wavelength bands and constrained the radius of the grains to be 0.2--0.7\,$\micron$ \citep{miotello2012_114-426}.

Atacama Large Millimeter/submillimeter Array (ALMA) observations of 114--426 detected the submillimeter continuum emission from large dust grains in the disk, while the submillimeter CO lines were seen in absorption against the nebular background \citep{mann2014_ALMA7ONC,bally2015_114-426}. The velocity profile of the CO measurement constrained the stellar mass to 0.4--1\,$M_\sun$, making this disk's large outer extent especially remarkable. 

In the present paper, we analyze James Webb Space Telescope (JWST) NIRCam images of the 114--426 disk in 12 wide-, medium-, and narrowband filters spanning 1--5\,$\micron$. These images show convincing evidence for the 3\,$\micron$ water ice absorption feature in both (1) lines of sight dominated by scattered light as well as (2) the outer ansae of the disk detected in silhouette against the bright background nebula. This analysis provides the first demonstration of NIRCam's ability to shed light on the nature of water ice in young circumstellar disks and provides some of the strongest direct constraints on the amount of water ice at large disk radii given the disk's unique orientation and position in the ONC.

\section{Methods}
\label{sec:methods}

\begin{figure*}
\epsscale{1.17}
\plotone{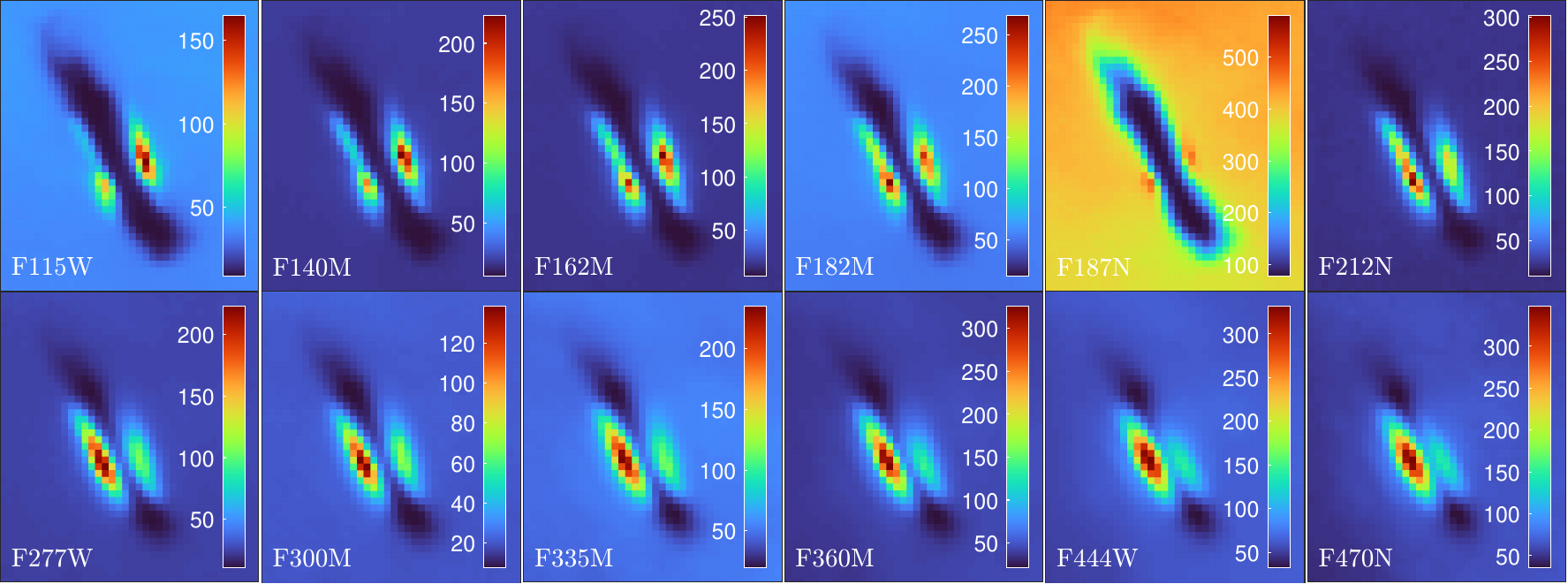}
\caption{Images of 114--426 in the 12 NIRCam bands, ordered from shortest to longest wavelength. The SW channel images have been resampled onto the coarser pixel scale of the LW channel images. Each panel is 2$\farcs$7$\times$2$\farcs$4 with north up and east left. The color bar shows the brightness in units of MJy sr$^{-1}$.}
\label{fig:gallery}
\end{figure*}

JWST/NIRCam imaged a large region ($10\farcm9 \times 7\farcm5$) of the ONC, including 114--426, for Cycle 1 GTO Program 1256 (PI: McCaughrean). For a complete overview of these observations see \citet{mccaughrean2023_ONCNIRCAM}. Twelve filters were used spanning the 1--5\,$\micron$ wavelength range: F115W, F140M, F162M, F182M, F187N, F212N, F277W, F300M, F335M, F360M, F444W, and F470N. The first six of these are from the short wavelength (SW) channel and the other six from the long wavelength (LW) channel.

We use the customized pipeline reduction of the images described by \citet{mccaughrean2023_ONCNIRCAM}. While the Level~2 pipeline-processed individual images available directly from the Mikulski Archive for Space Telescopes database yielded reasonable results, the default Level~3 processing to combine the many images into large-scale mosaics did not. In addition to astrometric errors, the default pipeline did not result in good matching of the bright and variable nebulosity across the Orion Nebula, leading to large-scale intensity steps and thus erroneous nebular fluxes. The custom pipeline processing corrected for the astrometric offsets and delivered smoothly matched large-scale mosaics.

For quantitative analysis, we crop the 12 images to smaller regions centered on 114--426. To align the images precisely, we oversample the images to a finer pixel scale. The native pixel size is 31 mas for the SW images and 63 mas for the LW images. We oversample the SW images onto a $4\times 4$ subgrid and the LW images onto an $8\times 8$ subgrid. We rotate the six SW images clockwise by 0.5$^\circ$ to correct the rotational offset between the SW and LW channels measured by \citet{mccaughrean2023_ONCNIRCAM}. We then apply small translational shifts at the subgrid level using the unsaturated point source located $\sim$3$\arcsec$ northeast of 114--426 as a guide, thereby bringing the 12 images into alignment. We then resample our images back to the coarser pixel scale of the LW channel before conducting the bulk of our analysis.

\begin{figure*}
\epsscale{1.17}
\plotone{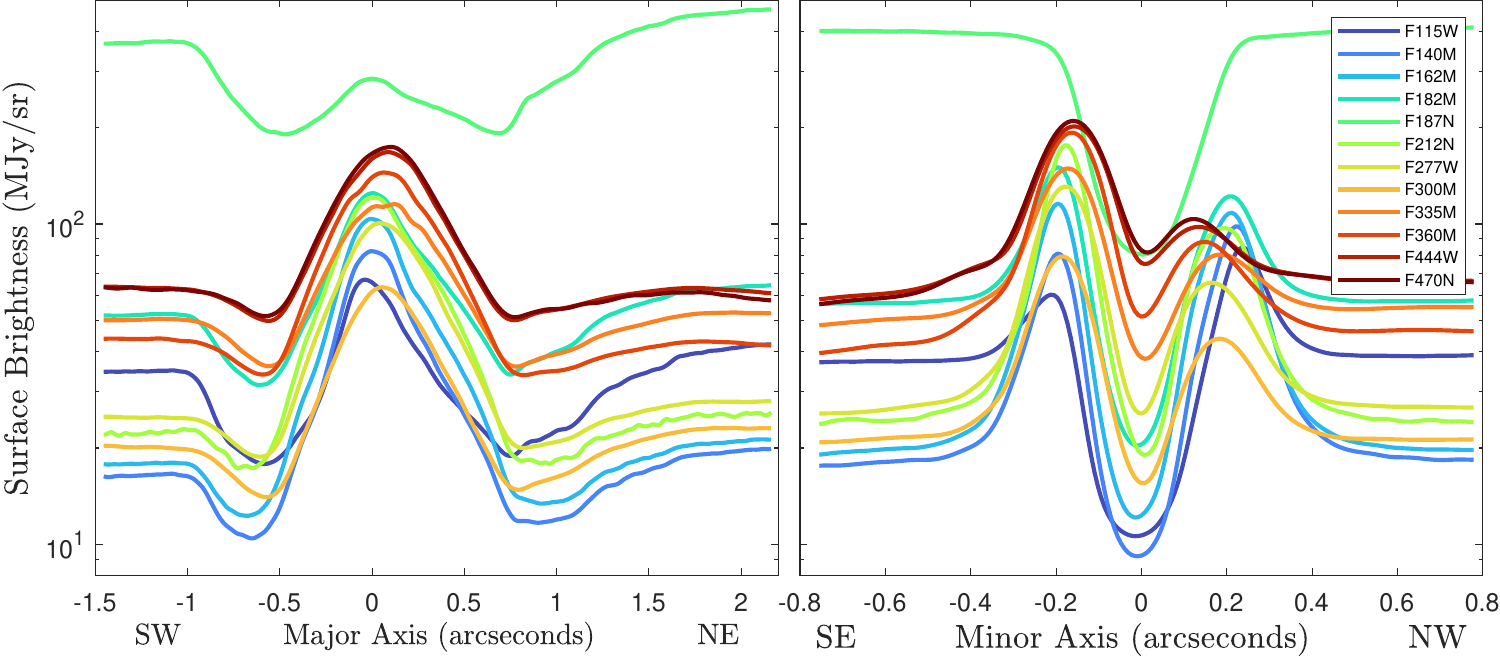}
\caption{Left: surface brightness cuts along major axis of the disk in a $0\farcs6$ wide rectangular aperture. Right: surface brightness cuts along the minor axis of the disk in a $1\farcs1$ wide rectangular aperture. The $y$-axis is the mean value of the pixels across the aperture. The cuts are extracted from the spatially oversampled images.}
\label{fig:cuts}
\end{figure*}

One factor that has complicated prior analyses of the 114--426 disk is that the telescope point-spread function (PSF) blurs bright emission from the surrounding nebula and the scattered light lobes into the silhouette, potentially leading to an overestimation of the transmitted flux \citep{mccaughrean1996AJ_silhouettes}. Because the diffraction-limited PSF size increases with wavelength, this could bias the observed slope of the absorption spectra. 

We quantify the magnitude of this effect by creating a second set of oversampled images, as above. This time, the images for the shortest 11 wavelengths are smoothed to bring them all to the resolution of the longest-wavelength and thus lowest-resolution image, F470N. To do so, we approximate the PSF of each band with a symmetric 2D Gaussian with FWHM equal to that of the empirical PSF according to the online documentation\footnote{https://jwst-docs.stsci.edu/jwst-near-infrared-camera/nircam-performance/nircam-point-spread-functions}. We convolve each image by a Gaussian with a standard deviation of the difference (in quadrature) between the PSF of the F470N band and that of the band being convolved. Finally, this set of images is again resampled back to the LW channel pixel scale. We analyze both the unconvolved and convolved sets of images, with some results from the latter presented in the Appendix.

We estimate the pixel-to-pixel noise in each image by computing the standard deviation of brightness values in a rectangular aperture offset from the disk. For the unconvolved images, these noise levels are 2.5\%, 3.0\%, 3.2\%, 2.8\%, 2.8\%, 3.7\%, 3.6\%, 4.5\%, 10.1\%, 6.8\%, 4.7\%, and 4.8\% of the median flux in the region in the 12 bands in order of increasing wavelength. When comparing the brightness between bands, we also include a 2\% flux calibration uncertainty. This is conservative, as most bands have $\le$2\% uncertainty, according to the current online documentation.

To study the disk in absorption (Section \ref{sec:SEDs}), we require an estimate of the brightness of the nebula behind the disk. To do so, we define an annular ellipse around the disk. The inner edge of this ellipse has a semi-major and semi-minor axis of 28 and 12 LW pixels, respectively, and the width of the ellipse is 16 pixels. The ellipse is rotated to align with the northeast-to-southwest orientation of the disk major axis. In most bands we notice a slight gradient in the brightness across the annulus, so we fit the brightness within the annulus with a plane. This plane serves as a model for the nebular background at any point in the image. To estimate the systematic uncertainty due to the specific annulus used, we repeat the calculation many times using annuli of different sizes. The variation in the results is much smaller than the pixel-to-pixel uncertainty, so our analysis is robust against the choice of annulus.

\section{Results and Analysis}
\label{sec:results}

\begin{figure*}
\epsscale{1.17}
\plotone{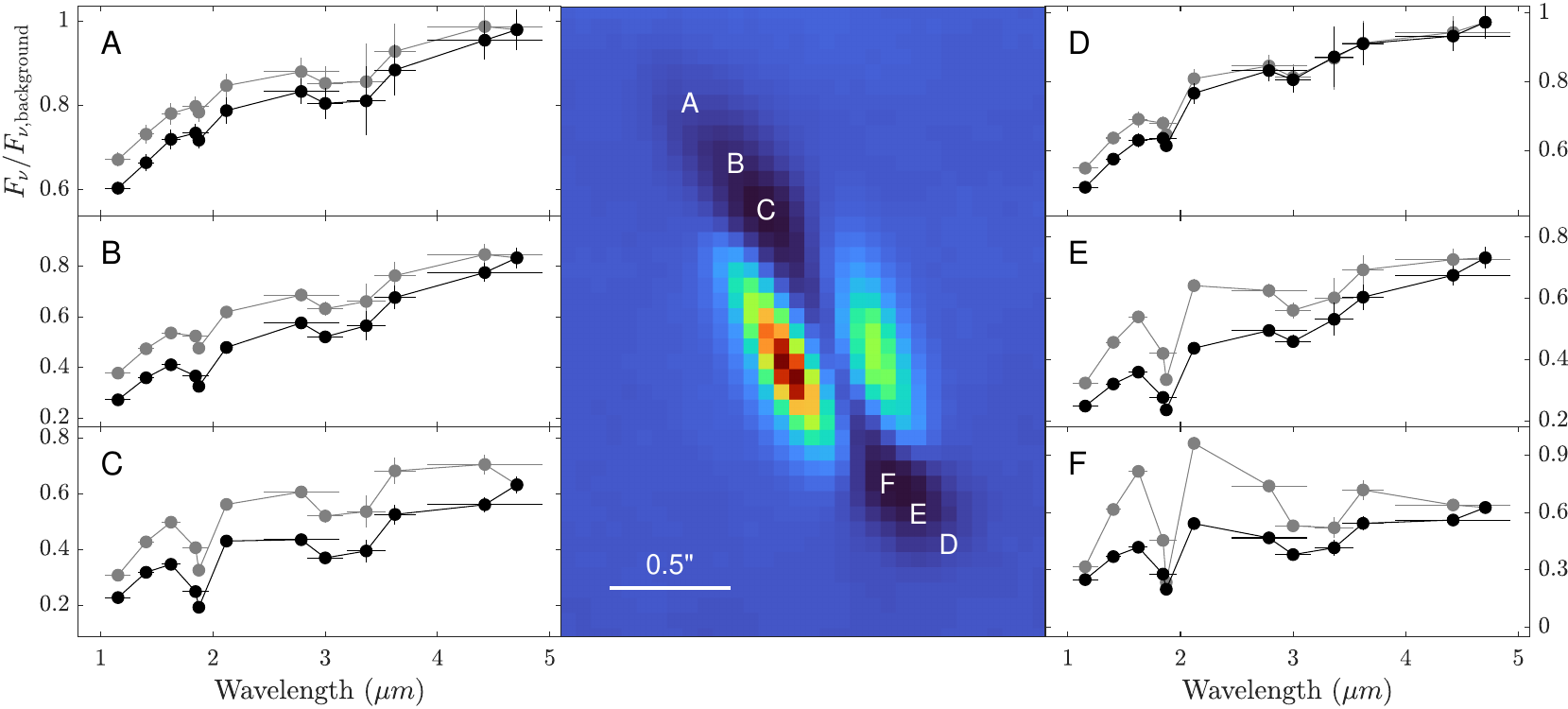}
\caption{Image of the disk in the F300M filter ($2\farcs6\times2\arcsec$) with north up and east left. Letters (A--F) mark six pixels in the outer ansae of the disk seen purely in silhouette against the background nebula. Absorption spectra at each of these pixels are shown in the panels to the left and right of the image. The spectra consist of measurements from the 12 NIRCam bands. Black spectra are extracted from the unconvolved images while gray spectra are extracted from the convolved images. Horizontal bars depict the bandwidths. All absorption spectra show a dip at 3\,$\micron$ due to the presence of water ice. Note that the range of the $y$-axis varies between panels.}
\label{fig:AbsSEDs}
\end{figure*}

\subsection{Disk Morphology}
\label{sec:morphology}
Figure~\ref{fig:highres} shows detailed images of 114--426. The left panel is a color composite of the four wide- and medium-width SW bands plus F277W, and the right panel shows the F187N band. In both images, the pixel scale is oversampled and the intensities are scaled logarithmically to compress the dynamic range and enhance fainter details. Figure~\ref{fig:gallery} displays a gallery of the 12 NIRCam images of 114--426, this time all sampled to the LW pixel scale for a more direct comparison and displayed with a linear brightness scale. Figure~\ref{fig:cuts} presents surface brightness cuts along the major and minor axes of the disk.

The images reveal a bilobed emission pattern surrounding a dark lane and flanked by even darker extensions seen in silhouette against the nebular background. The dark ansae curve away from the midplane in opposite directions, giving the disk a warped S-shaped structure. The ends of the ansae appear wispy with the northeast silhouette being more extended and having a hook-like shape. To first order, this morphology is consistent with that seen in HST images \citep{mccaughrean1998_114-426,miotello2012_114-426}.

The F187N image differs from the other bands. This narrowband filter is centered on the Pa$\alpha$ emission line at 1.875$\micron$, thus admitting significant light from the background \mbox{H\,{\footnotesize II}} and photodissociation region (PDR) while minimizing any continuum contribution. The scattered light lobes are barely visible, and the whole radial and vertical extent of the disk appears in silhouette. A similar appearance is evident in HST optical images that trace the H$\alpha$ line \citep{mccaughrean1998_114-426}, not least because the diffraction limit of the images is very similar (the 2.7$\times$ larger diameter of the JWST primary mirror offsetting the 2.9$\times$ longer wavelength of the Pa$\alpha$ line.)  

A detailed examination of the scattered light lobes shows a wavelength-dependent brightness asymmetry between them. At wavelengths shorter than 2\,$\micron$, the northwest lobe is brighter, while at wavelengths longer than 2\,$\micron$ the asymmetry flips and the southeast lobe is brighter. This behavior was seen with HST/NICMOS when comparing images at 0.57, 1.1, 1.6, and 2.0\,$\micron$ \citep{mccaughrean1998_114-426}, and here we find that the trend continues out to 4.7\,$\micron$. The pattern is clear from the redder color of the southeast lobe in the left panel of Figure~\ref{fig:highres} and from the minor axis cuts shown in Figure~\ref{fig:cuts}. A similar behavior is seen in edge-on disks around 2MASS J16281370-2431391 \citep[aka The Flying Saucer;][]{grosso2003_flyingsaucer} and IRAS 04302+2247 \citep[aka The Butterfly Star;][]{villenave2024_butterfly}, although in the latter case the brightness flip occurs at longer wavelengths between 12.8 and 21.1\,$\micron$. 

\begin{figure*}
\epsscale{1.17}
\plotone{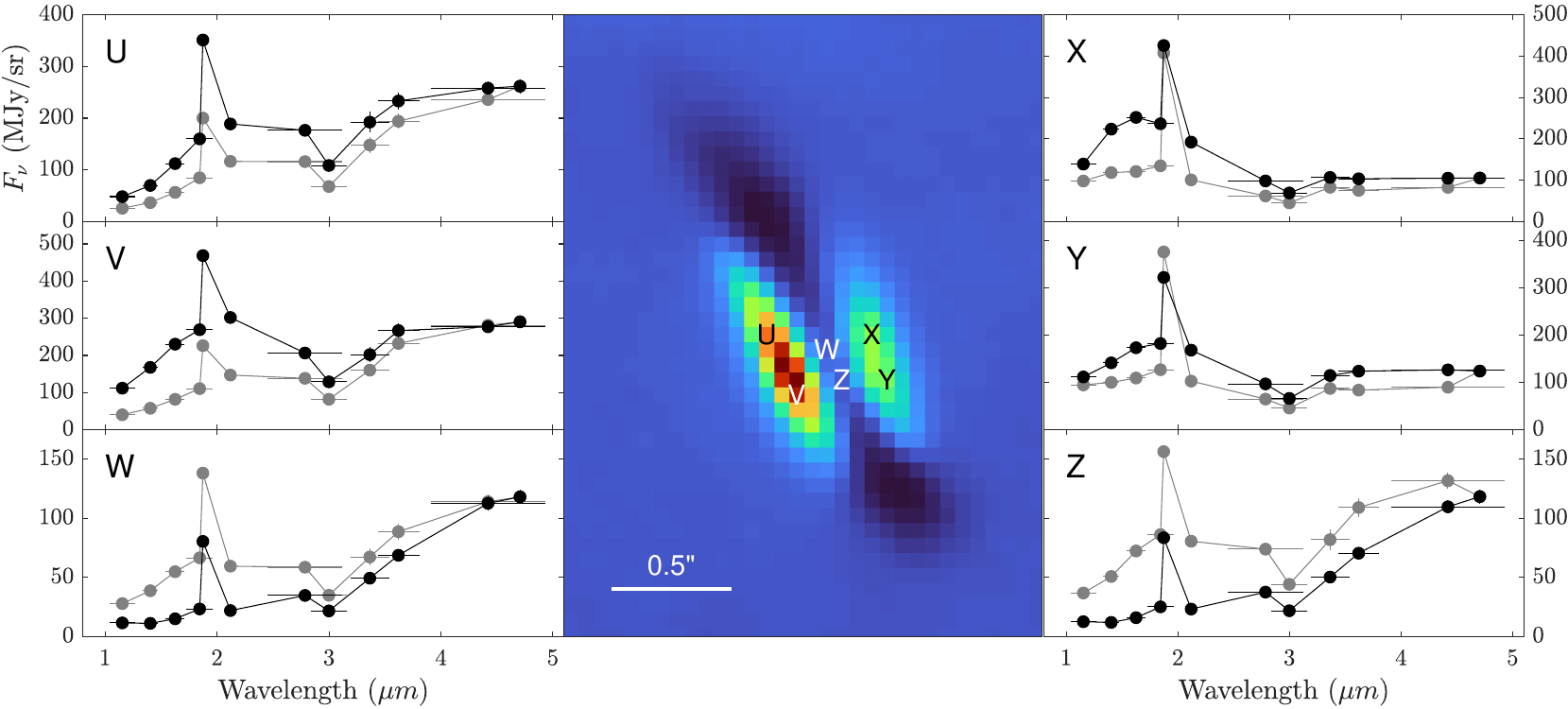}
\caption{Image of the disk in the F300M filter ($2\farcs6\times2\arcsec$) with north up and east left. Letters (U--Z) mark six pixels where the signal is dominated by the two lobes of scattered light and the dark lane between them. Spectra at each of these pixels are shown in the panels to the left and right of the image. The spectra consist of measurements from the 12 NIRCam bands. Black spectra are extracted from the unconvolved images while gray spectra are extracted from the convolved images. Horizontal bars depict the bandwidths. All spectra show a dip at 3\,$\micron$ due to the presence of water ice. Note that the range of the $y$-axis varies between panels.}
\label{fig:DiskSEDs}
\end{figure*}

There is also a lateral asymmetry where the peak brightnesses of the two lobes are offset from each other along the disk major axis. Such lateral asymmetries are found in the majority of imaged edge-on disks \citep{villenave2024_butterfly}. Radiative transfer models show that both wavelength-dependent brightness asymmetries and lateral asymmetries can arise when there is a tilted inner disk component \citep{villenave2024_butterfly}. Thus, the scattered light morphology suggests there is a tilted inner disk in 114--426, perhaps due to a binary star or massive planetary companion.

The geometry of the dark lane also varies with wavelength. The radial extent of the silhouette ansae appears to shrink with increasing wavelength, as does the width of the dark lane between the two scattered light lobes. Part of this trend can be explained by the diffraction-limited PSF of JWST growing with wavelength, but the trend is still evident in the images after convolving them to a common resolution (see the Appendix). Physically, this behavior is expected from edge-on disks when the dust opacity decreases with increasing wavelength \citep{watson2007_PPV,duchene2024_Tau042021}. However, we notice a deviation from this trend in the F300M image. In this band the dark lane is wider than in the F277W and F335M images that bracket it in wavelength. This indicates an increase in opacity at 3\,$\micron$. We explore the spectral behavior of the disk and the likely origin of this additional opacity in more detail in the following sections.

\subsection{Disk Spectra}
\label{sec:SEDs}

\subsubsection{Silhouette Region}
\label{sec:silhouettespectra}

Here we examine the spectral behavior at various locations in the 114--426 disk. We begin with the ansae of the dark lane seen in silhouette against emission from the background nebula. Figure~\ref{fig:AbsSEDs} shows the absorption spectra from six pixels in the silhouette. The points are labeled A--F and their locations are indicated on the image in the center of the figure. The absorption spectra are computed by dividing the brightness of the pixel by the brightness of the background at that location, derived as discussed in Section~\ref{sec:methods}. Taking this ratio normalizes the inherent brightness variation with wavelength of the background nebula. It also cancels out any systematic band-to-band calibration uncertainty. The error bars on this ratio do include the statistical uncertainty computed in Section~\ref{sec:methods}. Horizontal bars on the plotted points indicate the bandwidths. Black spectra are extracted from the unconvolved images while gray spectra are extracted from the convolved images, and we discuss the difference between these in the Appendix. 

At the shortest three wavelengths (F115W, F140M, and F162M) the disk becomes increasingly transparent with increasing wavelengths. This is the expected behavior of small ($\lesssim$ few micrometer) dust grains and continues the trend seen from shorter wavelengths with HST \citep{miotello2012_114-426}. We quantify the grain sizes in Section~\ref{sec:iceabundance}. At the darkest parts of the silhouette, 20\%--40\% of the background light is transmitted at these wavelengths. 

\begin{figure*}
\epsscale{1.15}
\plotone{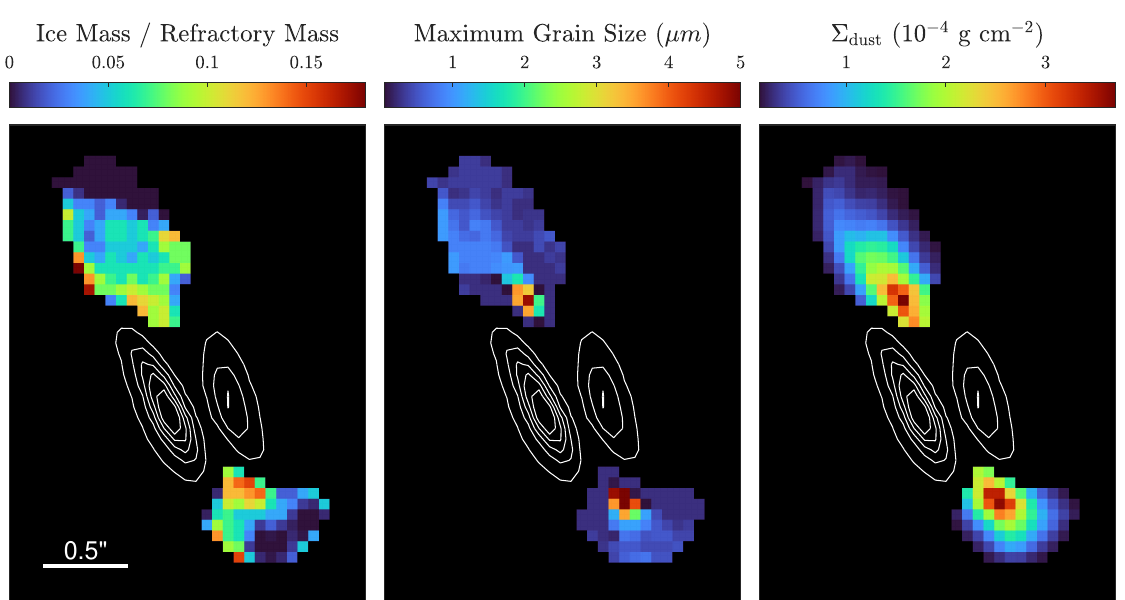}
\caption{Best-fit ice-to-refractory mass ratio (left), maximum grain size (center), and surface density (right) in each pixel of the silhouette of the disk. White contours shown the scattered light lobes of the disk in the F300M filter. We do not fit pixels in this region or in the dark lane between the lobes.}
\label{fig:fitimages}
\end{figure*}

The next two bands (F182M and F187N) show a dip in the absorption spectra at many of the pixels. These bands include the Pa$\alpha$ line, in which the background is very bright. The fact that the disk absorbs more photons at this wavelength than at adjacent wavelengths is unexpected and cannot be explained by the opacity of bare or icy dust grains. This suggests that the gas in the disk contributes to the opacity in these bands. We discuss this possibility more in Section~\ref{sec:Paalpha}.     

Another prominent dip in the absorption spectra is seen in the F300M band. This filter has a medium bandwidth and does not sample a bright line from the nebular background. Thus, we associate the feature with water ice in the disk, as seen in many other protostellar and protoplanetary systems (see Section~\ref{sec:introduction}.) The 3\,$\micron$ water ice feature is relatively broad, but its shape is not fully sampled by the available photometry. The F300M band does not probe the full depth of the feature, and the F277W and F335M bands likely sample a mix of the wings of the feature and the neighboring continuum. 

At the longest wavelengths (F360M, F444W, and F470N) the absorption spectra continue to rise, with the disk becoming nearly totally transparent at the outer edge of the ansae (e.g., locations A and D).

\subsubsection{Scattered Light Region}
\label{sec:scatteredspectra}

In Figure \ref{fig:DiskSEDs} we show the spectra at various locations centered on the scattered light lobes (U, V, Y, and Z) and in the dark lane between them (W and X). Because the disk here is not seen in absorption relative to the background, we do not divide the brightness by the background. The error bars include the measured statistical uncertainty plus the 2\% calibration uncertainty. As with Figure \ref{fig:AbsSEDs}, the black spectra are extracted from the unconvolved images while the gray spectra are extracted from the convolved images, and the difference is discussed in the Appendix.

In these locations, the Pa$\alpha$ line is evident in emission from an excess in the F187N band. The amplitude does not trace the true intensity of the line because it is diluted by the width of the band, which is likely wider than that of the line. Interpreting the source of these Pa$\alpha$ photons is difficult. They could arise from the background nebula and are partially transmitted through this region of the disk, although unlike in the silhouette, the optical depth of the scattered light lobes cannot easily be quantified. Other potential sources of these Pa$\alpha$ photons are accretion onto the central star with the photons then scattering out of the disk or emission from excited hydrogen gas in the disk. Future spectroscopic studies covering additional nebular lines could test the degree to which the lines in the scattered light lobes resemble those from the background. Spectra of the scattered light lobes would also be useful to constrain the spectral type of the central star. 

A dip in the spectra from water ice at 3\,$\micron$ is evident in this region of the disk as well. This is consistent with predictions from radiative transfer models of ice-bearing edge-on disks \citep{ballering2021_diskices,sturm2023_HH48modelII} and from observations of such disks \citep[e.g., ][]{sturm2023_HH48ERS}. The brightness of the disk in the F300M filter -- even in the center of the dark lane -- does not go to zero. This is consistent with measurements of the 3\,$\micron$ feature in the dark lane of other disks such as HH48\,NE \citep{sturm2023_HH48ERS}. One possible explanation is that photons from ice-free regions of the disk scatter into the same lines of sight as those from icy regions. This adds a floor of continuum at 3\,$\micron$, preventing the flux from dropping to zero even if the disk midplane is actually very ice rich. It could also be due to the NIRCam PSF blurring signal into the midplane from the bright lobes. Furthermore, PAH emission from the disk at 3.3 $\micron$ could influence the shape of the feature, as was observed in the spectrum of the edge-on disk HH48\,NE \citep{sturm2023_HH48ERS,sturm2024_HH48MIRI}. Interpreting the details of the spectra in the scattered light region requires fitting the data with radiative transfer models, which is beyond the scope of this study.

\subsection{Quantifying the Water Ice Abundance, Grain Sizes, and Dust Mass}
\label{sec:iceabundance}

\begin{figure*}
\epsscale{1.17}
\plotone{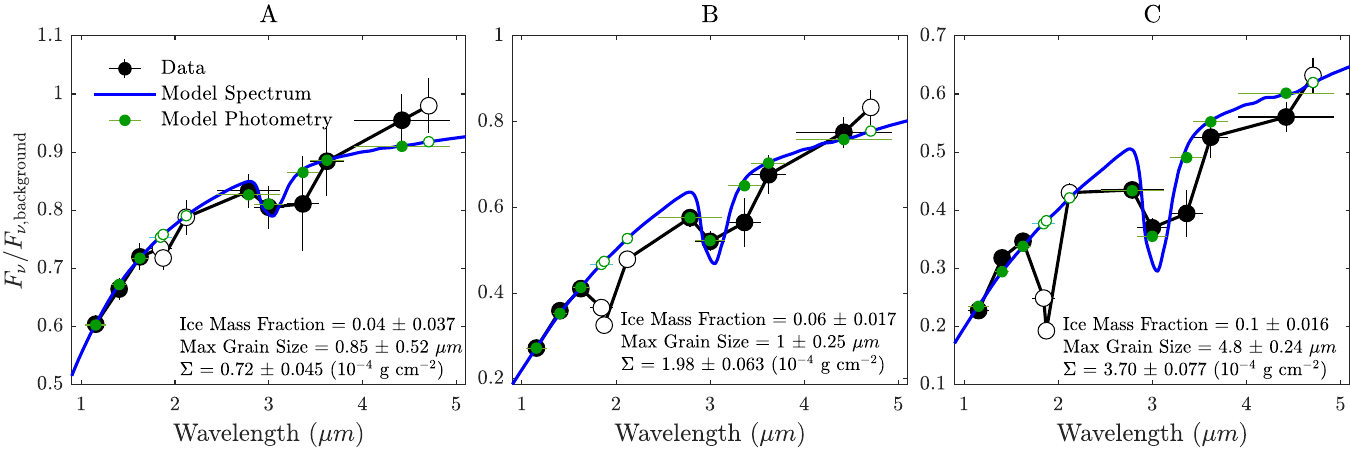}
\caption{Model fits to the absorption spectra at three locations (A, B, and C) as noted in Figure \ref{fig:AbsSEDs}. Black points are the data, the blue curve is the model at high spectral resolution, and the green points are the synthetic photometry of the model. Open points are not included in the fit. The best-fit model parameters are noted in the lower right of each panel.}
\label{fig:fitspectra}
\end{figure*}

We now return our attention to the silhouette of the disk. Given these lines of sight primarily intercept material at the very edges of the disk under more similar physical and chemical conditions compared to lines of sight directly toward the central star, a relatively straightforward approach can be used to extract the water-ice-to-dust composition as described in the following paragraphs.

The ratio of the measured brightness over the background brightness is related to the amount of absorbing material by
\begin{equation}
F_\nu/F_{\nu\text{,background}} = e^{-\kappa_{\rm abs}\Sigma_{\rm dust}}, 
\label{eq:model}
\end{equation} 
where $\kappa_{\rm abs}$ is the dust absorption opacity (units of cm$^2$~g$^{-1}$) and $\Sigma_{\rm dust}$ is the dust surface density (units of g~cm$^{-2}$). $\Sigma_{\rm dust}$ reflects the amount of dust along the line of sight and is independent of wavelength, while $\kappa_{\rm abs}$ varies with wavelength and is set by the grain size and composition. We choose to include only the absorption opacity and not the scattering opacity because scattering will be preferentially in the forward direction and thus photons from the background will mostly continue into our line of sight. The model assumes there is no emission from the silhouette of the disk or from nebular (or interstellar) material between the disk and Earth. This is supported by the conclusion that silhouette disks like 114--426 likely reside in the foreground of the Orion Nebula \citep{mccaughrean1996AJ_silhouettes,miotello2012_114-426}.  

\begin{figure*}
\epsscale{1.17}
\plotone{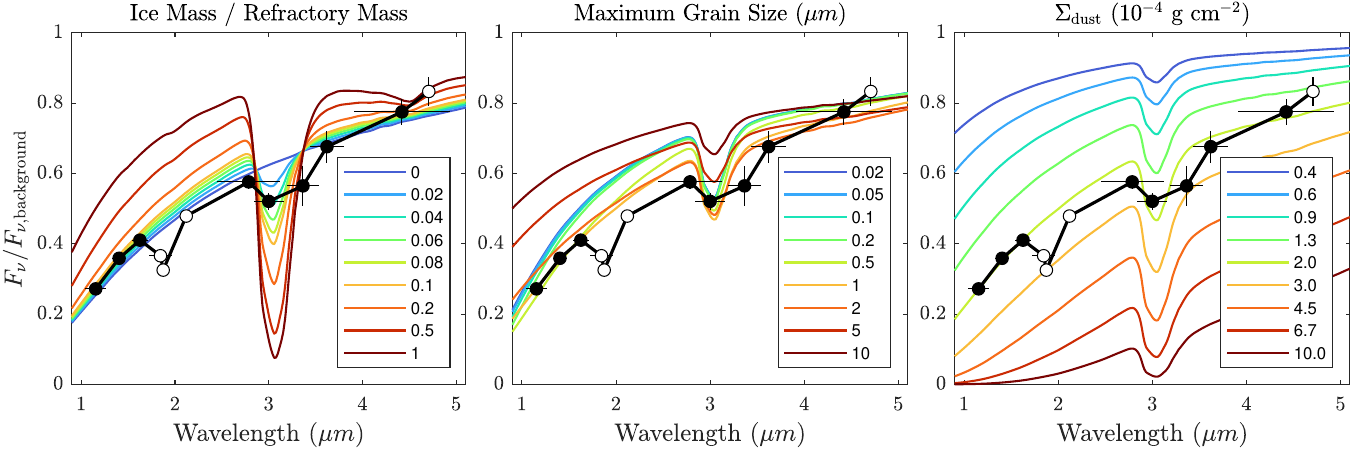}
\caption{Demonstration of how varying the ice-to-refractory mass ratio (left), maximum grain size (middle), and dust surface density (right) changes the model absorption spectrum. The black points are the measurements at point~B from Figures~\ref{fig:AbsSEDs} and~\ref{fig:fitspectra}. In each panel, the two model parameters not being varied are fixed at the best-fit values to these data.}
\label{fig:parademo}
\end{figure*}

We generate model opacity spectra using the code \texttt{optool} \citep{dominik2021_optool}. The minimum grain size is fixed at 0.005 $\micron$, while the maximum grain size is allowed to vary as a free parameter and the size distribution is a power law with an index of $-3.5$ as is typically assumed. The grains are modeled as a refractory core with an ice mantle. The core composition is set to a mixture of 87\% amorphous pyroxene (with a 70\% Mg content) and 13\% amorphous carbon by mass. This is a commonly assumed core composition \citep{woitke2016_DIANA} and a default option in  \texttt{optool}. The mass of the water ice mantle is parameterized as a fraction of the core mass. The optical constants for pyroxene, carbon, and water ice are from \citet{dorschner1995}, \citet{zubko1996_carbon}, and \citet{warren2008_waterice}, respectively. Note that optical constants of ice vary by up to 15\% depending on the temperature and porosity \citep[e.g.,][]{rocha2024_waterice}.

There are thus three free parameters in our model: (1) the ice-to-refractory mass ratio, (2) the maximum grain size, and (3) $\Sigma_{\rm dust}$. The first two parameters enter Equation \ref{eq:model} through $\kappa$. We generate model absorption spectra on a finely sampled wavelength grid then compute synthetic photometry for the 12 NIRCam bands by integrating the model weighted by the band transmission profiles. We do not include all bands in the fit. We exclude the F182M and F187N bands because there is apparent excess absorption of Pa$\alpha$ photons in these bands, which is not included in our simplified model. We also exclude the other two narrowband filters in the set: F212N and F470N. These are centered on molecular H$_2$ transitions and thus could potentially be contaminated by absorbing gas in the disk. We use a grid search to find the model that best fits the remaining eight bands by minimizing the $\chi^2$ metric
\begin{equation}
\chi^2 = \sum\frac{\left(\text{Data}\,F_\nu/F_{\nu\text{,back}} - \text{Model}\,F_\nu/F_{\nu\text{,back}}\right)^2} { \sigma(F_\nu/F_{\nu\text{,back}})^2},
\end{equation}
where the sum is over bands included in the fit. We fit each pixel in the silhouette of the disk independently. We fit pixels where the absorption depth in the shortest wavelength band is $\leq$0.9 and where the scattered light does not contaminate the absorption spectrum. We also fit these models to the spatially convolved images, and the results are discussed in the Appendix. 

Maps of the best-fit parameters are shown in Figure \ref{fig:fitimages}. The ice-to-refractory mass ratio (left panel) varies from 0 to 0.18. While there is considerable scatter, the ice fraction tends to decrease with increasing disk radius. The maximum grain size (middle panel) decreases radially outward as well as upward away from the disk midplane. We find maximum grain sizes of $\sim$0.25--1\,$\micron$ in the outer parts and up to 5\,$\micron$ at smaller radii (but note that 5\,$\micron$ is the largest value we considered in the fitting). The grain sizes are further discussed in Section \ref{sec:grainsizes}. The mass surface density (right panel) decreases radially, as expected. Values range from 0.13--3.7 $\times 10^{-4}$ g cm$^{-2}$ in the silhouette.

By summing the surface density from each pixel, we can derive the total mass of ice and dust in the silhouettes. At a distance of 390\,pc, each pixel corresponds to a physical area of $1.35 \times 10^{29}$ cm$^2$. We find 0.27\,$M_\oplus$ of dust and ice in the northeast ansa and 0.19\,$M_\oplus$ in the southwest ansa, totaling 0.46 $M_\oplus$ of solid material in the silhouette. \citet{bally2015_114-426} find a dust mass of 9.9\,$M_\oplus$ in the scattered light region of the disk from ALMA continuum measurements. So while the silhouette occupies a sizeable fraction of the disk's area on the sky, it harbors a smaller fraction of the total mass because it lacks larger millimeter-size grains where most of the solid mass is sequestered.

We show a spectral fit at three pixels in Figure \ref{fig:fitspectra}. These pixels are at the locations labeled A, B, and C in Figure \ref{fig:AbsSEDs}. The values of the best-fit model parameters, along with their uncertainties, are noted in the figure. These locations illustrate a range of ice mass ratios seen across the northeast ansa. The quality of these fits is typical for the pixels throughout the silhouette. The model fits the rising absorption spectrum in the shortest three bands well. The model greatly overpredicts the F182M and F187N bands (excluded from the fit) that may be impacted by preferential absorption of Pa$\alpha$ photons. The model also sometimes overpredicts the data in the excluded F212N band. The F277W and F300M points are fit well. F300M samples the core of the ice absorption feature (although the photometry does not trace the full depth of the feature seen in the high-resolution model), while F277W and F335M are affected by the wings of the feature. F335M is regularly overpredicted by the model. The final two bands that we fit, F360M and F444W, show varying degrees of agreement with the model. Overall this suggests that the shape of the red edge of the water feature is not reproduced correctly in our model, or that there are other sources of opacity that we have not considered. We discuss this further in Section \ref{sec:rededge}.

Figure \ref{fig:parademo} demonstrates the impact of each of the three free parameters on the model behavior. The depth of the 3\,$\micron$ feature is highly sensitive to even small water ice fractions (left panel). High ice fractions also shift the model vertically at wavelengths outside of the feature, but such shifts are likely degenerate with the other model parameters. Increasing the maximum grain size flattens the slope of the model in the 1--3 \,$\micron$ region (middle panel). The variation is most drastic between sizes of 0.2 and 5\,$\micron$ with very little variation among the models at smaller sizes. At sizes larger than 5\,$\micron$ the slope remains constant but the models shift vertically. This is degenerate with other model parameters, which is why we limit the fits to maximum grain sizes $\leq$5\,$\micron$. Varying the solid surface density scales the entire absorption spectrum vertically, as expected (right panel).

\section{DISCUSSION}
\label{sec:discussion}

\subsection{Red Edge of the Water Ice Feature}
\label{sec:rededge}

The ice absorption models we present in Section \ref{sec:iceabundance} result in 3\,$\micron$ features that do not extend to long enough wavelengths to reproduce the depth seen in the F335M band (Figure \ref{fig:fitspectra}). An extended red edge to this feature is also routinely observed in spectra toward protostars and through background clouds \citep[e.g.,][]{smith1989_3umiceprotostars,dartois2002_icesprotostars,noble2013_AkariIces}, including most recently with JWST/NIRSpec \citep{mcclure2023_iceageclouds}. \citet{dartois2024_icegrainsizeJWST} demonstrated how to successfully fit the extended red edge of these NIRSpec spectra. Their model is more sophisticated than ours in three ways: (1) they include additional ice species beyond water, e.g., CH$_3$OH and NH$_3$, (2) they derive the dust optical properties using the discrete dipole approximation to account for more realistic grain shapes, (3) they use the RADMC-3D code \citep{dullemond2012_RADMC3D} to model the radiation transfer of the background photons through the cloud. Such procedures are beyond the scope of this work and not warranted given the coarse spectral information in the photometric data. However, if future observations of 114--426 at higher spectral resolution detect the extended red edge, these more sophisticated modeling approaches would be justified. 

\subsection{Origins of the Pa$\alpha$ Absorption}
\label{sec:Paalpha}

As described in Section \ref{sec:SEDs}, the absorption spectra in the silhouette of the disk show dips in the F182M and F187N filters. Given that the \mbox{H\,{\footnotesize II}} and PDR nebular background is bright in atomic hydrogen lines (including Pa$\alpha$), and that we are unaware of any plausible dust or ice opacity feature at this wavelength, it seems most likely that the dip in the absorption spectra in these bands is due to preferential absorption of Pa$\alpha$ photons. 

Pa$\alpha$ absorption is a transition from the $n$=3 to $n$=4 electronic states of atomic hydrogen. Is it plausible that the disk hosts a reservoir of hydrogen in the $n$=3 state? It is unlikely that gas in the outer disk is hot enough to populate the $n$=3 level thermally. The gas may be excited from the ground state by the absorption of Ly$\beta$ photons. These could arise from the central star via its chromosphere or accretion shocks. Alternatively, Ly$\beta$ photons could arise from massive stars in the ONC or diffuse nebular radiation. 114--426 does not display the cometary-shaped ionization front surrounding the disk that is typical of photoevaporating ``proplyd" disks seen throughout the ONC, meaning this system is shielded from most ionizing extreme-ultraviolet photons. However, \citet{miotello2012_114-426} argued that 114--426 does receive a significant dose of external far-ultraviolet (FUV) radiation, which includes the Ly$\beta$ line. 

As the focus of this work is analyzing the water ice, we defer detailed calculations of the state of the gas in the outer parts of 114--426 to future studies.

\subsection{Grain Growth in the Outer Disk}
\label{sec:grainsizes}

Our finding of grains with radii up to $\sim$1\, $\micron$ in the outer regions of the disk confirms that dust has grown larger than the sizes found in the interstellar medium (ISM). This maximum size is also consistent with ALMA images of 114--426, which found that the submillimeter continuum (tracing even larger grains) only extends throughout the scattered light region of the disk and not into the silhouette \citep{mann2014_ALMA7ONC,bally2015_114-426}. The lack of pebbles in the outer regions could be because (1) grain growth beyond a few microns is inefficient at large disk radii, (2) the large grains that formed in this region have spiraled inward due to aerodynamic drag \citep{weidenschilling1977_drift}, or (3) the outer ansae represent outflowing material in which only small grains are entrained \citep{miotello2012_114-426}.

The grain sizes we derive differ compared to the results of some prior studies of the 114--426 silhouette. \citet{throop2001_114-426} found no difference in absorption between H$\alpha$ (0.656 $\micron$) and Pa$\alpha$ images, and so concluded that grains had grown to sizes larger than 5 $\micron$. \citet{shuping2003_114-426} extended this analysis by detecting somewhat less absorption in a ground-based Br$\alpha$ (4.05 $\micron$) image, and concluded the grains were 1.9--4 $\micron$ in size. Our finding of somewhat smaller grains is likely due to our use of continuum bands (and exclusion of the Pa$\alpha$ band) when deriving grain sizes, whereas these prior studies exclusively used images in narrow filters centered on the atomic hydrogen lines to maximize contrast of the silhouettes against the background nebula. As shown in Figures \ref{fig:fitspectra} and \ref{fig:parademo}, our models fit a rise in the absorption spectra from 1 to 2 $\micron$ and ignore the drop in the F187N filter. Models relying on the F187N filter would not fit this rise, inferring a flatter absorption spectrum and thus larger grains. \citet{miotello2012_114-426} fit the absorption in five HST filters, four of which had wide bandwidths, and found 0.2--0.7\,$\micron$ grains in the northeast silhouette. Our results agree with those findings while using data out to longer wavelengths.

\subsection{Survivability of Water Ice}
\label{icesurvival}

The ONC hosts luminous OB stars that bathe the other cluster members and their disks with a strong radiation field. 114--426 resides at a projected separation of 0.2\,pc from the massive star $\theta^1$ Ori C in the Trapezium region, but the line-of-sight separation is unknown so the true separation may be significantly larger. \citet{miotello2012_114-426} estimate that 114--426 experiences an external FUV flux of $\sim 2 \times 10^3 G_0$, although this too is very uncertain. $G_0$ is defined as 10$^8$ photons cm$^{-2}$ s$^{-1}$ and is approximately equal to the strength of the external FUV field in the local ISM around the sun \citep{habing1968_G0}. 114--426 is likely exposed to a level of external irradiation significantly higher than experienced by disks in low-mass star-forming regions (e.g., Taurus, Lupus, and Chamaeleon) but lower than experienced by Trapezium Cluster members closer to the center of the Orion Nebula often associated with ionized proplyds ($\geq\!10^5\,G_0$). The key question to answer is whether it is reasonable to find water ice in the outer parts of a disk in this intermediate environment.

One impact of an external radiation field is that it may heat the disk above the sublimation temperature of volatile species, keeping them in the gas phase. \citet{haworth2021_warmdust} investigated this possibility and found that the water snow line remains largely immune from external heating if the disk resides more than 0.02\,pc from $\theta^1$\,Ori C, which is the case for 114--426. Their models found the midplane temperature in the outer disk of a system like 114--426 to be 50--60\,K. This value is below the sublimation temperature of water (150\,K), comparable to the sublimation temperature of CO$_2$ (50\,K), and above the sublimation temperature of CO (20\,K). ALMA detected CO gas from 114--426 in absorption against the background nebula \citep{bally2015_114-426}. The background CO has a brightness temperature of 50--70\,K, and the optically thick CO in the disk was seen to be 12--27\,K below this level. Thus, observations of the gas temperature are consistent with the disk being cold enough to maintain water ice against thermal desorption.

\begin{figure}
\epsscale{1.17}
\plotone{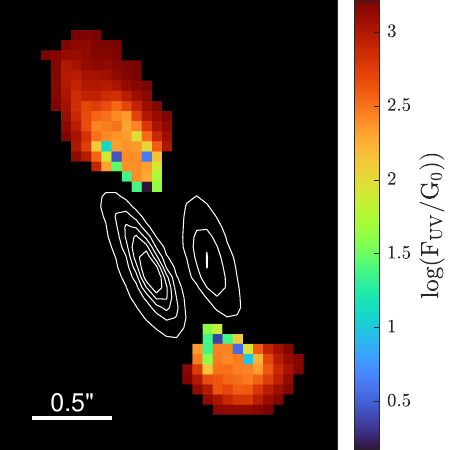}
\caption{FUV field strength in the disk calculated assuming an incident field of $2 \times 10^3\,G_0$ and attenuation in the disk based on the dust properties and column density found by fitting the NIRCam data. White contours show the scattered light lobes of the disk in the F300M band.}
\label{fig:FUV}
\end{figure}

More importantly, the external UV field may nonthermally photodesorb water molecules off of the grains and into the gas phase. The efficiency of this process depends on the strength of the FUV field within the icy regions of the disk. To make an order of magnitude estimate, we adopt the incident FUV flux from \citet{miotello2012_114-426} and quantify the extinction arising from the disk material itself, $F_{\rm UV} = 2 \times 10^3\,G_0\,e^{-0.5\kappa_{\rm abs}\Sigma_{\rm dust}}$. The factor of 0.5 arises because we are interested in the attenuation of the external field halfway through the total column density. This simplification does not take into account photons entering from varying angles \citep{cleeves2013_CR}, but it provides a reasonable estimate assuming most photons enter from the direction of the background nebula. Here $\kappa_{\rm abs}$ is the absorption opacity of the dust in the FUV. We compute the opacity spectrum from 912 to 2000\AA{} with the \texttt{optool} code using the dust properties found when fitting to the near-IR data in Section~\ref{sec:iceabundance}. We average the opacity spectrum over this wavelength range, weighted by a 39,000\,K blackbody (the effective temperature of $\theta^1$\,Ori C) to approximate the spectral shape of the external field in the core of the Orion Nebula. Figure \ref{fig:FUV} shows the predicted UV field at each position within the silhouette of the disk. The outer less-attenuated (and less icy) radii have $F_{\rm UV}$ spanning 100--1000\,$G_0$. In the inner, denser regions, $F_{\rm UV}$ drops to 10--100\,$G_0$. 

Given this FUV field strength, do we expect most of the water to be in ice or gas phase? Motivated by the formalisms presented by \citet{hollenbach2009_H2O} and \citet{oberg2009_photodesorptionII}, we can estimate the equilibrium amount of water ice versus vapor by equating the photodesorption rate (per volume) of ice into gas with the adsorption rate (per volume) of gas into ice 
\begin{equation}
    F_{\rm UV} Y (1-e^{-x/l}) n_{\rm gr} \sigma_{\rm gr} = n_{\rm H_2O,gas} v_{\rm H_2O} n_{\rm gr} \sigma_{\rm gr}.
    \label{eq:equilibrium}
\end{equation}
Note that this is a simplified calculation in that we neglect any formation or destruction of water molecules (ice or gas), and we neglect other desorption processes. $n_{\rm H_2O,gas}$ is the volume density of water in the gas phase, $n_{\rm gr}$ is the volume density of dust grains, and $\sigma_{\rm gr}$ is the cross section of the average individual grain. 

\begin{figure*}
\epsscale{1.17}
\plotone{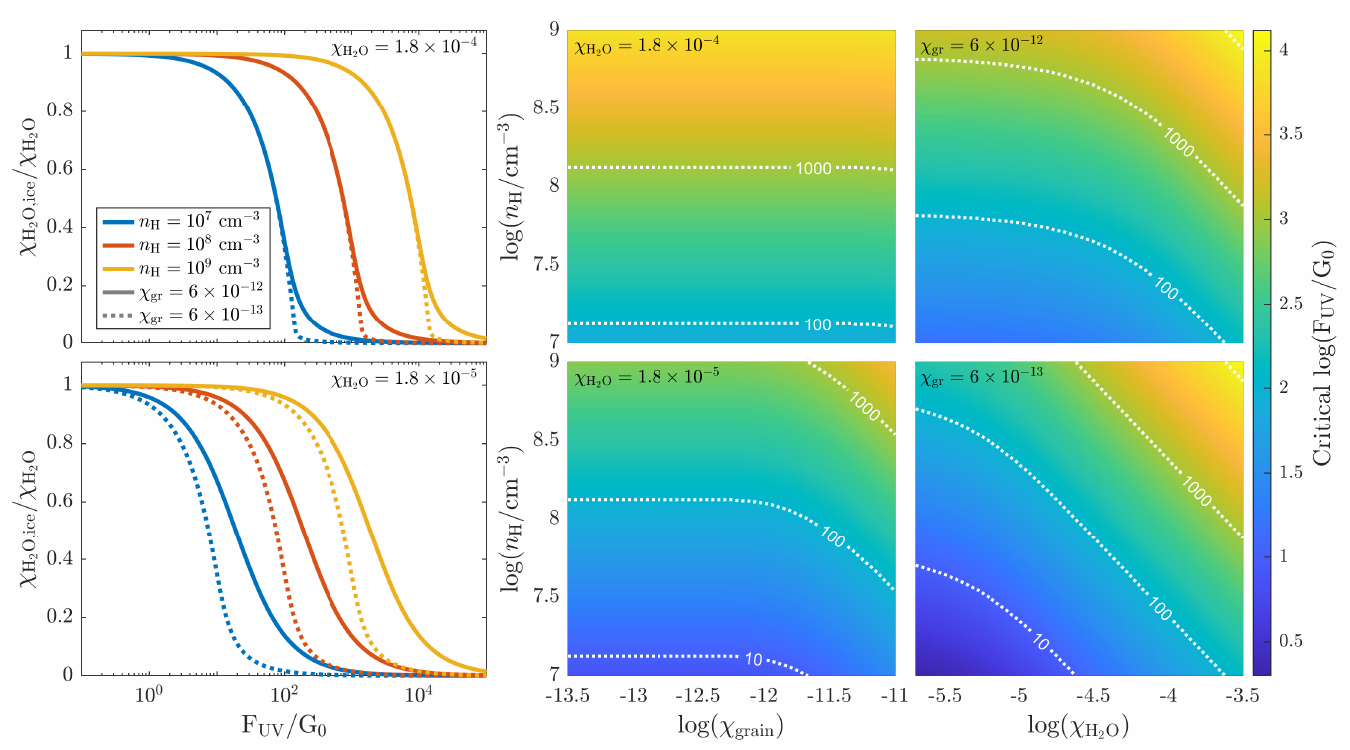}
\caption{Left: predicted fraction of water remaining in the ice phase as a function of FUV field strength. The top panel shows the case with a fiducial total water abundance of $\chi_{\rm H_2O}$ = 1.8$\times$10$^{-4}$ and the bottom panel when the water abundance is reduced by an order of magnitude. The three line colors are for different gas densities ($n_{\rm H}$). The solid lines use a fiducial value for the dust grain abundance ($\chi_{\rm gr} = 6\times10^{-12}$) while the dashed lines show a grain abundance reduced by an order of magnitude. Right: critical value of $F_{\rm UV}$ (where half the water is in the ice phase) as a function of $n_{\rm H}$ and $\chi_{\rm gr}$ (for two fixed water abundances), and as a function of $n_{\rm H}$ and $\chi_{\rm H_2O}$ (for two fixed grain abundances). The results assume equilibrium between adsorption and photodesorption of water ice and are derived by solving Equation~\ref{eq:equilibrium} numerically.}
\label{fig:photodesorption}
\end{figure*}

It is convenient to express each volume density in the form of an abundance ($\chi$) relative to the hydrogen volume density of the bulk gas ($n_{\rm H}$), i.e., $n_{\rm H_2O,gas}$ = $\chi_{\rm H_2O,gas}~n_{\rm H}$, $n_{\rm gr}$ = $\chi_{\rm gr}~n_{\rm H}$, etc. We are ultimately interested in the fraction of the total water locked in ice, i.e., $\chi_{\rm H_2O,ice}/\chi_{\rm H_2O}$. We assume a fiducial value for the total water abundance $\chi_{\rm H_2O}$ = $\chi_{\rm H_2O,gas}$ + $\chi_{\rm H_2O,ice}$ = 1.8$\times$10$^{-4}$. In reality, this could be reduced by chemical processing, UV-driven photodissociation, or if ice is sequestered into larger grains and transported to the inner disk. The bulk gas density of the outer parts of the 114--426 disk is not known. \citet{bally2015_114-426} estimated $n_{\rm H}$ $>$ 10$^7$ cm$^{-3}$, so we explore the results with values of $n_{\rm H} = 10^7$--$10^9$ cm$^{-3}$.

$v_{\rm H_2O}$ is the thermal speed of water molecules in the gas, given by
\begin{equation}
v_{\rm H_2O} = \sqrt{\frac{8 k_B T_{\rm gas}}{\pi m_{\rm H_2O}}}. 
\end{equation}
Here $k_B$ is the Boltzmann constant, $m_{\rm H_2O}$ = 3$\times$10$^{-23}$ g is the mass of a water molecule, and $T_{\rm gas}$ is the gas temperature, which we assume to be 40\,K given the ALMA CO observations by \citet{bally2015_114-426} discussed above. These values result in a water thermal speed of $v_{\rm H_2O} = 216$~m\,s$^{-1}$

$Y$ is the photodesorption yield, for which we adopt a value of $2.6 \times 10^{-3}$ water molecules liberated per FUV photon, following Equation 10 of \citet{oberg2009_photodesorptionII} assuming a dust temperature of 40\,K\@. $x$ is the number of monolayers of ice on each grain (or if $x<1$, the fraction of the grain surface coated in ice). It is related to the abundance of ice according to
\begin{equation}
x = \frac{\chi_{\rm H_2O,ice}}{\chi_{\rm gr} N_{\rm sites}}
\label{eq:fs}
\end{equation}
where $N_{\rm sites} = 1\times 10^6$ is the number of water molecules needed to coat a grain. We adopt a fiducial value of $\chi_{\rm gr} = 6\times10^{-12}$ for the grain abundance, which is appropriate for a gas-to-dust ratio of 100. In reality, $\chi_{\rm gr}$ could be reduced below the fiducial value in the outer disk due to the growth and inward drift of grains, or it could be raised slightly if gas (but not dust) is lost from the outer disk in a wind \citep{throop2005_evapplanetesimals}. $l$ is the diffusion length of the ice, which in effect is the depth into the ice mantle affected by photodesorption. We assume a value of $l$ = 3 monolayers.

By combining the above results we can solve Equation~\ref{eq:equilibrium} numerically for the fraction of water in the ice phase. However, before discussing the results, it is worth pointing out two limiting cases that can be solved analytically. First, when ice is abundant enough to fully coat the grains with many monolayers ($x/l \gg 1$) the term $(1-e^{-x/l}) \approx 1$. Then $n_{\rm gr}$ and $\sigma_{\rm gr}$ drop out of Equation \ref{eq:equilibrium} because both processes depend linearly on the total grain surface area available. In this case we find
\begin{equation}\label{eq:abund1}
    \frac{\chi_{\rm H_2O,ice}}{\chi_{\rm H_2O}} = 1 - \frac{F_{\rm UV}Y}{\chi_{\rm H_2O} n_{\rm H}  v_{\rm H_2O}}.
\end{equation}
The second case is when the ice abundance is low, $x/l \ll 1$, and $(1-e^{-x/l}) \approx x/l$. Here the total water abundance drops out of the equation and we find 
\begin{equation}\label{eq:abund2}
    \frac{\chi_{\rm H_2O,ice}}{\chi_{\rm H_2O}} = \left[1 + \frac{F_{\rm UV}Y}{\chi_{\rm gr} n_{\rm H} v_{\rm H_2O} N_{\rm sites} l} \right]^{-1}.
\end{equation}

We plot the results of this analysis in Figure~\ref{fig:photodesorption}. The left panels shows the fraction of water in the ice phase versus $F_{\rm UV}$ for three values of $n_{\rm H}$ (colors), with fiducial $\chi_{\rm H_2O}$ (top) and reduced $\chi_{\rm H_2O}$ (bottom), and with fiducial $\chi_{\rm gr}$ (solid) or reduced $\chi_{\rm gr}$ (dashed). The right four panels of Figure \ref{fig:photodesorption} explore the effect of the model parameters more fully. They map out the ``critical" $F_{\rm UV}$, at which half the water is in the ice phase, for a range of $n_{\rm H}$, $\chi_{\rm gr}$, and $\chi_{\rm H_2O}$ combinations. 

There are many significant unknowns regarding the physical and chemical conditions of the 114--426 outer disk, leading to values of the critical $F_{\rm UV}$ ranging from $<$10 to 10,000 $G_0$ in the parameter space we explore. However, we find that for fiducial values of $\chi_{\rm gr}$ and $\chi_{\rm H_2O}$, and gas density $\geq$10$^8$ cm$^{-3}$, the majority of the water can remain as ice even in the outer parts of the disk where the incident FUV field is unattenuated. Thus, we conclude that it is entirely plausible for water ice to survive against the effects of UV photodesorption in this environment, and the detection of water ice is not at odds with what we know about this disk.

If, however, future studies of this disk suggest a lower $n_{\rm H}$ or $\chi_{\rm H_2O}$, which would make ice survival less plausible in these conditions, there are a couple of possibilities that may explain the observations. First, the strength of the incident FUV field on the disk may have been overestimated by \citet{miotello2012_114-426}. Second, the dark ansae may not represent bound icy grains on stable orbits but instead material leaving the disk in a photoevaporative wind \citep{miotello2012_114-426}. In this scenario, the material may have only recently reached this outer exposed region, and the ice may not yet have had time to desorb.

The ability of water ice to survive in externally irradiated disks like 114--426 is significant because most stars, including the Sun, form in cluster environments like the ONC \citep{lada2003,winter2022_photoevapreview}, and an external FUV flux of 10$^2$--10$^4$ $G_0$ is the most typical field strength experienced by young stars \citep{fatuzzo2008_UVfields}. Two other edge-on disks further from the center of the Orion Nebula (132--1832 and 216--0939) have 3\,$\micron$ water ice detections \citep{terada2012_iceOrion,terada2012_iceOrion2}. With the addition of water ice detected in 114--426, there is growing evidence for water ice in intermediately irradiated regions of clustered star formation -- the typical environment of planet formation in our Galaxy.

\section{Summary}
\label{sec:summary}

We present an analysis of the 114--426 edge-on silhouette disk in 12 JWST NIRCam images. 114--426 is large on the sky and resides in front of the background emission of the Orion Nebula. This offers a unique opportunity to study the disk material in absorption. The NIRCam images span the prominent 3\,$\micron$ water ice absorption feature, allowing us to search for this important volatile in a reservoir of planet-forming material. Our  main findings include the following.
\begin{itemize}
    \item The two lobes of scattered light above and below the disk midplane are laterally offset from each other and exhibit a brightness asymmetry that flips with wavelength. These phenomena suggest there is a tilted inner disk in this system.
    \item The spectra of the scattered light lobes and the dark lane between them show a dip at 3\,$\micron$ due to water ice. The abundance of ice in these regions cannot be directly inferred from the depth of the feature due to scattering effects.
    \item The outer ansae of the disk, seen in silhouette against the background nebula, also show a dip in their absorption spectra at 3\,$\micron$ indicating the presence of water ice.
    \item We fit a model absorption spectrum of water-ice-coated grains to each pixel in the silhouette. The ice-to-refractory mass ratio ranges from 0 to 0.18, the maximum grain size ranges from 0.25 to 5\,$\micron$, and the total mass of solids in this region is 0.46\,$M_\oplus$. Observations at higher spectral resolution would improve the accuracy of these findings.
    \item We also observe a dip in the absorption spectra at the wavelength of Pa$\alpha$, suggesting there is excited atomic hydrogen in the absorbing material. The gas may be excited by UV photons from the central star or external sources, but a quantitative assessment will require future study.   
    \item The balance between adsorption and UV photodesorption indicates that some water can remain in the ice phase in this environment, in agreement with the observations. However, this balance depends on several properties of this system that are currently uncertain.    
    \item This study provides direct observational evidence that water ice can survive in an intermediately irradiated disk, the typical planet-forming environment in the Galaxy.
\end{itemize}

Higher spectral resolution near-IR observations of this unique system would confirm many of the results presented here and open the door to further discoveries. The NIRSpec instrument on board JWST, for instance, has an integral field unit mode that would be ideally suited for this task. At higher resolution the water ice absorption feature could be better isolated from the neighboring continuum and its spectral profile measured, enabling a more accurate determination of the ice abundance. The shape of the profile can indicate whether the water ice is amorphous or crystalline \citep{mastrapa2009_waterice}. An extended red edge to the feature, hinted at in the NIRCam data, can also reveal if the ice is pure or mixed with other species such as CH$_3$OH and NH$_3$ \citep{dartois2024_icegrainsizeJWST}. Furthermore, such observations could search for the distinct narrower absorption features from more volatile CO$_2$ and CO ice at 4.27 and 4.67\,$\micron$. Finally, higher spectral resolution observations would be ideal to detect absorption (or emission) from gas phase atomic and molecular species in the silhouette, notably the Pa$\alpha$ line seen in this study.

\medskip
We thank the referee for their feedback that improved the clarity of the paper. This work is based on observations made with the NASA/ESA/CSA James Webb Space Telescope. The data used are available in the Barbara A. Mikulski Archive for Space Telescopes at the Space Telescope Science Institute, which is operated by the Association of Universities for Research in Astronomy, Inc., under NASA contract NAS 5-03127 for JWST\@. The dataset ID is 10.17909/vjys-x251. 

N.P.B. acknowledges support from the Virginia Initiative on Cosmic Origins (VICO), NSF grant no.\ AST-2205698, SOFIA Award 09-0181, and NASA grant no.\ 21-XRP21-0016. 
L.I.C. acknowledges support from the David and Lucille Packard Foundation, Research Corporation for Science Advancement Cottrell Fellowship, NASA ATP 80NSSC20K0529, and NSF grant no.\ AST-2205698.
R.D.B. acknowledges support from the Virginia Initiative on Cosmic Origins (VICO) and NSF grant no.\ AST-2206437.
S.G.P. acknowledges support through the ESA research fellowship program at ESA's ESTEC, and would like to thank Victor See for helpful discussions and Katja Fahrion for valuable insights on the JWST calibration pipeline.

The data presented in this paper were obtained with the Near Infrared Camera (NIRCam) on the NASA/ESA/CSA James Webb Space Telescope under Cycle~1 program 1256, as part of the Guaranteed Time Observation allocation made to M.J.M. upon selection as one of two ESA Interdisciplinary Scientists on the JWST Science Working Group (SWG) in response to NASA AO-01-OSS-05 issued in 2001. He would like to thank STScI instrument scientists and program reviewers Elizabeth Nance, Massimo Robberto, and Mario Gennaro, and also Tony Roman for help in scheduling and execution.


\facilities{JWST}

\bibliographystyle{aasjournal}
\bibliography{NPB}

\appendix

\section{Results after image convolution}
\label{sec:convoresults}

In the body of the paper, we analyzed the 12 images after they were interpolated, aligned, and resampled to the common LW pixel scale. Here we briefly discuss the equivalent results and analysis after also degrading the resolution of the 11 shorter-wavelength images to match that of the longest-wavelength F470N image, as discussed in Section~\ref{sec:methods}. 

\begin{figure*}
\epsscale{1.17}
\plotone{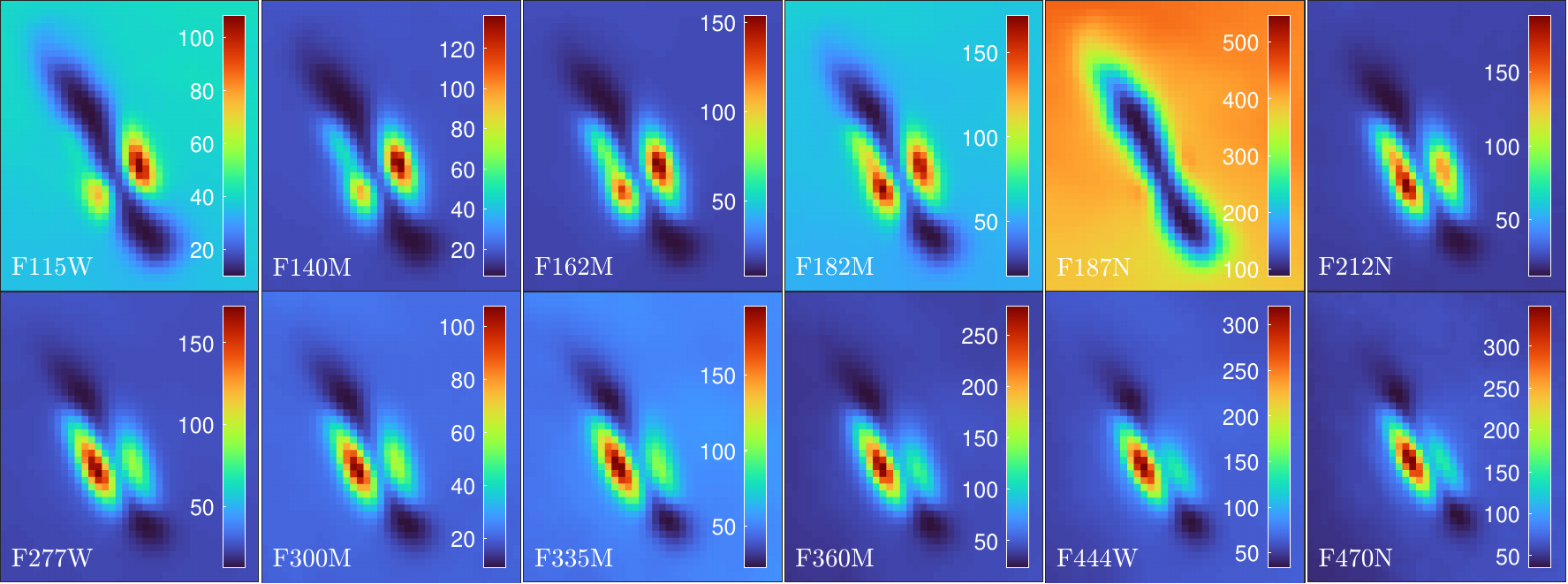}
\caption{Similar to Figure \ref{fig:gallery}, except the images are convolved to a common diffraction-limited spatial resolution, namely that of the F470N image.}
\label{fig:galleryconv}
\end{figure*}

Figure \ref{fig:galleryconv} shows a gallery of images convolved to the F470N resolution. Differences are minimal with respect to the corresponding unconvolved images. The wavelength-dependent brightness asymmetry and lateral offset between the two scattered light lobes remain apparent. The shrinking of the dark midplane with increasing wavelength is also still evident, although the trend is somewhat counteracted by the higher degree of smoothing in the shorter-wavelength images.

The spectra extracted at certain pixels from the convolved images are shown in gray in Figures \ref{fig:AbsSEDs} and \ref{fig:DiskSEDs}. They reveal approximately the same pattern as the unconvolved spectra (shown in black) -- including the Pa$\alpha$ and water ice features -- except they are shifted to slightly higher or lower values. All of the absorption spectra in the silhouette are shifted up, indicating less absorption. This is expected, as more flux from the surrounding nebula is blurred into the silhouette. The spectra from the scattered light lobes (U, V, X, and Y) are fainter after convolution because some of the flux is blurred away. The spectra from the midplane (W and Z) are slightly brighter because additional flux from the scattered light lobes is blurred into this region.

\begin{figure*}
\epsscale{1.15}
\plotone{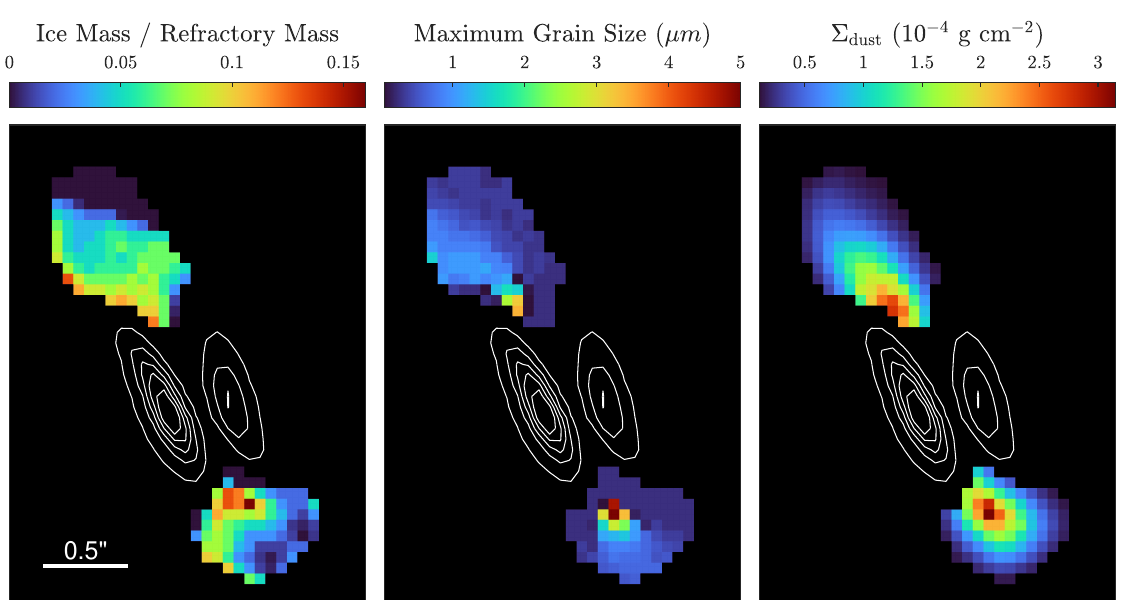}
\caption{Similar to Figure \ref{fig:fitimages}, except the model absorption spectra are fit to the data after convolving the images to the F470N diffraction-limited spatial resolution.}
\label{fig:fitimagesconv}
\end{figure*}

We fit the absorption spectra of all pixels in the silhouette of the convolved images using the same method as for the unconvolved images (Section \ref{sec:iceabundance}), and maps of the best-fit parameters are shown in Figure \ref{fig:fitimagesconv}. We find the ice-to-refractory mass ratio is generally higher by up to  $\sim$50\%, the maximum grain size agrees to within $\sim$40\%, and the solid surface density is $\sim$10\%--30\% lower when using the convolved images compared to the unconvolved images. Overall, we conclude that the variation in spatial resolution across the NIRCam bands used for this analysis does not significantly bias the results presented in the main text.

\end{document}